\documentclass[letter]{aa}

\usepackage{twoopt}
\bibpunct{(}{)}{;}{a}{}{,}             
\makeatletter
  \newcommandtwoopt{\citeads}[3][][]{\href{http://adsabs.harvard.edu/abs/#3}%
    {\def\hyper@linkstart##1##2{}%
     \let\hyper@linkend\@empty\citealp[#1][#2]{#3}}}
  \newcommandtwoopt{\citepads}[3][][]{\href{http://adsabs.harvard.edu/abs/#3}%
    {\def\hyper@linkstart##1##2{}%
     \let\hyper@linkend\@empty\citep[#1][#2]{#3}}}
  \newcommandtwoopt{\citetads}[3][][]{\href{http://adsabs.harvard.edu/abs/#3}%
    {\def\hyper@linkstart##1##2{}%
     \let\hyper@linkend\@empty\citet[#1][#2]{#3}}}
  \newcommandtwoopt{\citeyearads}[3][][]%
    {\href{http://adsabs.harvard.edu/abs/#3}
    {\def\hyper@linkstart##1##2{}%
     \let\hyper@linkend\@empty\citeyear[#1][#2]{#3}}}
\makeatother

\makeatletter
\renewcommand*\aa@pageof{, page \thepage{} of \pageref*{LastPage}}
\makeatother

\usepackage{footnote}
\usepackage{amsmath}
\usepackage{amssymb}
\usepackage{xcolor}
\usepackage{graphicx}
\usepackage{pdflscape}
\usepackage{url}
\usepackage[varg]{txfonts}
\usepackage{bm}         
\usepackage{booktabs}

\newcommand{\JJ}{J0217$-$0820}

\DeclareGraphicsExtensions{.pdf,.png,.jpg}

\begin{document}

\title{ALMACAL XI: Over-densities as signposts for proto-clusters?}
\subtitle{A cautionary tale}

\author{
Jianhang~Chen\inst{\ref{inst1}}\thanks{Email: Jianhang.Chen@eso.org; cjhastro@gmail.com}\fnmsep
\and
R.\,J.~Ivison\inst{\ref{inst1}} \inst{\ref{inst2}} \inst{\ref{inst3}} \inst{\ref{inst4}} \inst{\ref{inst5}}
\and
Martin~A.~Zwaan\inst{\ref{inst1}}
\and
Anne~Klitsch\inst{\ref{inst6}}
\and
C\'eline P\'eroux\inst{\ref{inst1}} \inst{\ref{inst7}}
\and
Christopher C.~Lovell\inst{\ref{inst8}} \inst{\ref{inst9}}
\and
Claudia del P.~Lagos\inst{\ref{inst5}} \inst{\ref{inst10}} \inst{\ref{inst11}}
\and
Andrew D.~Biggs\inst{\ref{inst12}} 
\and
Victoria Bollo\inst{\ref{inst1}}
}

\institute{
European Southern Observatory (ESO), Karl-Schwarzschild-Strasse~2, D-85748 Garching, Germany\label{inst1}
\and
Department of Physics and Astronomy, Macquarie University, North Ryde, New South Wales, Australia\label{inst2}
\and
School of Cosmic Physics, Dublin Institute for Advanced Studies, 31 Fitzwilliam Place, Dublin D02 XF86, Ireland\label{inst3}
\and
Institute for Astronomy, University of Edinburgh, Royal Observatory, Blackford Hill, Edinburgh EH9 3HJ, UK\label{inst4}
\and
ARC Centre of Excellence for All Sky Astrophysics in 3 Dimensions (ASTRO 3D)\label{inst5}
\and
DARK, Niels Bohr Institute, University of Copenhagen, Jagtvej 128, 2200 Copenhagen, Denmark\label{inst6}
\and
Aix Marseille Universit\'e, CNRS, LAM (Laboratoire d’Astrophysique de Marseille) UMR 7326, 13388, Marseille, France\label{inst7}
\and 
Institute of Cosmology and Gravitation, University of Portsmouth, Burnaby Road, Portsmouth PO1 3FX, UK\label{inst8}
\and
Astronomy Centre, University of Sussex, Falmer, Brighton BN1 9QH, UK\label{inst9}
\and 
International Centre for Radio Astronomy Research (ICRAR), M468, University of Western Australia, 35 Stirling Hwy, Crawley, WA 6009, Australia\label{inst10}
\and 
Cosmic Dawn Center (DAWN), Denmark\label{inst11}
\and
UK Astronomy Technology Centre, Royal Observatory, Blackford Hill, Edinburgh EH9 3HJ, UK\label{inst12}
}

\date{Received 6 June 2023 / Accepted 29 June 2023}        

\titlerunning{ALMACAL XI --- a cautionary tale}
\authorrunning{Jianhang Chen et al.}

\abstract
{It may be unsurprising that the most common approach to finding proto-clusters is to search for over-densities of galaxies. 
Upgrades to submillimetre (submm) interferometers and the advent of the {\it James Webb Space Telescope} will soon offer the opportunity to find more distant candidate proto-clusters in deep sky surveys without any spectroscopic confirmation.
In this letter, we report the serendipitous discovery of an extremely dense region centred on the blazar, J0217$-$0820, at $z=0.6$ in the ALMACAL sky survey. 
Its density is eight times higher than that predicted by blind submm surveys.
Among the seven submm-bright galaxies, three are as bright as conventional single-dish submm galaxies, with $S_{\rm 870\mu m}\!>\!3$\,mJy. 
The over-density is thus comparable to the densest known and confirmed proto-cluster cores. 
However, their spectra betray a wide range of redshifts. 
We investigate the likelihood of line-of-sight projection effects using light cones from cosmological simulations, finding that the deeper we search, the higher the chance that we will suffer from such projection effects.
The extreme over-density around J0217$-$0820 demonstrates the strong cosmic variance we may encounter in the deep submm surveys. Thus,
we should also question the fidelity of galaxy proto-cluster candidates selected via over-densities of galaxies, where the negative $K$ correction eases the detection of dusty galaxies along an extraordinarily extended line of sight.
}


\maketitle

\section{Introduction}
\label{sec:introduction} 

Cosmological models predict the hierarchical evolution of structure
across cosmic time. Proto-clusters of galaxies are supposedly the
earliest over-densities to take shape, to form stars and (one way or
another) to light up. They should serve as excellent tracers of early
structure formation, as they are believed to evolve into the most massive
clusters in today's Universe \citep[see reviews in][]{Overzier2016,
  Alberts2022}.

For the few examples known, their extreme over-densities and active
ongoing star formation are consistent with model predictions for
dense nodes at intersections of the so-called `cosmic web'
\citep{Bond1996}. Their activity is indicative of feeding by cold
streams of gas from this web, which somehow sustains vigorous star
formation and nurses the growth of super-massive black holes
\citep[e.g.][]{Dekel2009}. Proto-clusters are thus ideal
laboratories for studying the interplay between baryons and various
feedback processes.

Traditionally (and for obvious reasons) statistical over-densities of galaxies have been used to identify proto-clusters. 
Pioneering studies have utilised broad-band photometry to
search for over-densities of Lyman-break galaxies \citep[LBGs, e.g.
][]{Steidel1998},  leading to the discovery of giant structures such
as that at $z=3.1$ in the SSA\,22 field.
Many more were found subsequently via deep narrow-band surveys,
searching for various strong line emitters, such as Ly\,$\alpha$ emitters
\citep[e.g.][]{Steidel2000, Matsuda2004, Ouchi2005} and H\,$\alpha$
emitters \citep[e.g.][]{Kurk2004a, Tanaka2011, Hayashi2012}. Large
multi-band photometric surveys have also inspired selection based on colour
\citep[e.g.][]{Zirm2008, Capak2011, Wylezalek2013a,
  Laporte2022}. Those surveys have preferentially targeted existing massive
galaxies or structures, such as high-redshift radio galaxies (HzRGs),
quasars, and giant Ly\,$\alpha$ blobs (LABs). Among them, HzRGs are
known to be hosted by massive galaxies \citep{Seymour2007} and
are therefore thought to trace the most massive dark matter halos.
Indeed, they have often been found to signpost over-dense regions and
are frequently embedded in proto-clusters \citep[e.g.][]{Stevens2003,
  Miley2008}. We also know that LABs are giant structures,
sometimes extending several hundred kpc from the central engine. They
are widely thought to be powered by starbursts and/or strong active
galactic nuclei (AGN) feedback, both preferentially triggered in
denser environments \citep[e.g.][]{Overzier2013, Umehata2019}.
Specifically, in the proto-typical proto-cluster, known as the Spiderweb at
$z=2.2$, all these methods have been used to reveal various galaxy
populations \citep{Kurk2000, Pentericci2002, Kurk2004a, Zirm2008,
  Hatch2011}.

Contemporaneously, advances in detector technology at submillimetre
(submm) wavelengths led to the discovery of submm galaxies
\citep[so called SMGs,][]{Smail1997, Hughes1998, Barger1998}. These galaxies
were initially selected at 850\,$\mu$m and were eventually found to be
dominated by distant $z>1$ dusty starburst galaxies, forming stars at
over $>\!100\,\text{M}_\odot\,\text{yr}^{-1}$, thereby revolutionising our
understanding of the most intense star-forming galaxies at high
redshift \citep[see the review by][]{Hodge2020}. Due to their extreme star formation rates (SFRs) and massive gas reservoirs,
SMGs are thought to trace peaks in the underlying density field and
are viewed as the most likely precursors of present-day ellipticals
\citep[e.g.][]{Lilly1999, Swinbank2006}, which has made them another
intriguing potential signpost for proto-clusters. Because of this,
searches for SMGs in known proto-cluster environments -- as well as
for galaxy over-densities around SMGs -- have been popular ways to
find proto-clusters and study their evolutionary status
\citep[e.g.][]{Dannerbauer2014, Umehata2018, Calvi2023}.

Recent advances in submm interferometry have accelerated the confirmation and 
resolved studies of the most extreme and rare proto-cluster cores.
Proto-cluster cores are supposed to be the densest known galaxy structures in
the early Universe, likely tracing the cores of the most massive dark
matter halos. These rare galaxy over-densities were first discovered
as single bright sources by single-dish telescopes with limited
spatial resolution, then later resolved into typically a dozen
starburst galaxies via deeper interferometric observations. In the
Distant Red Core (DRC) at $z=4.0$, for example, first identified via
imaging with \textit{Herschel}, the James Clerk Maxwell Telescope (JCMT)
and APEX by \citet{Ivison2016a}, 12 SMGs were later confirmed in the
central 300\,kpc region, along with four more gas-rich galaxies in the
outskirts \citep[e.g.][]{Oteo2018, Ivison2020}; 
in SPT\,2349$-$56, at $z=4.3$, 
14 dusty star-forming galaxies were found gathered within a 130-kpc
region \citep[][]{Miller2018}. These recent observational
breakthroughs suggest that we may be witnessing the most extreme
evolutionary stage of a proto-cluster, where the central region is
being transformed into something typical of today's clusters: a
massive central cD-type galaxy, such as NGC\,1275 in the Perseus
cluster. With the advent of the {\it James Webb Space Telescope (JWST)}, we
are likely to find many more such systems at higher redshifts
\citep[e.g.][]{Laporte2022, Jin2023, Morishita2023, Helton2023}.

However, the number of confirmed proto-clusters remains limited. 
Due to the expense of the spectroscopic follow-up required to confirm the
redshifts of member galaxies, searching for projected over-densities remains the most commonly used technique for identifying proto-clusters \citep[e.g.][]{Lammers2022}.
Because they can cover larger areas than interferometers, 
single-dish submm telescopes continue to target promising
candidates: for instance, HzRGs, LABs, as well as
over-densities found at other wavelengths, searching for additional SMGs
\citep[e.g.][]{Robson2014, MacKenzie2017, Zeballos2018, Li2020,
  Wang2021, Nowotka2022, Zhang2022, Li2023}.
Over-densities have been reported in many of
these systems, although the densities are only marginally
higher than those of the control fields and further follow-up often
gives mixed results.
For instance, \citet{Wylezalek2013} used far-IR/submm spectral energy distributions (SEDs) to constrain the
redshifts of galaxies towards the $z=3.8$ radio galaxy, 4C\,41.17,
finding just one of the many neighbouring far-IR galaxies to be at
similar redshift as 4C\,41.17, with most of the sources in
the foreground.
Similar false confirmations have also been reported in several proto-cluster candidates \citep{Chapman2015,Meyer2022}.

In this letter, we report the discovery of an extreme over-density
within the ALMACAL survey: seven point-like sources around an ALMA
calibration source, namely, the blazar, ICRF J021702.6$-$082052, at $z=0.6$ (hereafter, \JJ){}. Six of these are shown to be dusty star-forming galaxies (DSFGs) based on multi-band submm photometry. The over-density is comparable to the most extreme known, confirmed proto-cluster cores.

First, we summarise the ALMACAL survey and the data used in this work, in
\S\ref{sec:data}. We describe the over-density of submm galaxies around \JJ{} in
\S\ref{sec:overdensity}.  We unveil the redshifts of those
galaxies in \S\ref{sec:redshifts}, then we discuss the conclusions and
implications in \S\ref{sec:discussion}, using lightcones made from recent cosmological simulations to
better understand the statistics.

\begin{figure*}
   \centering
   \resizebox{1.0\hsize}{!}{\includegraphics{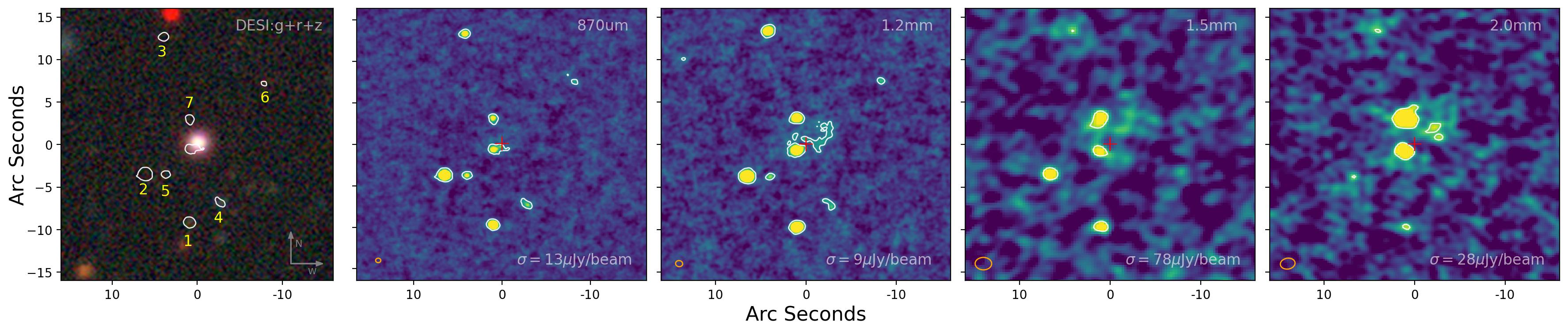}}
   \caption{Multi-wavelength images of \JJ. Left to right: RGB images from the DESI Legacy Imaging Surveys; ALMA images at 870\,$\mu$m, 1.2\,mm, 1.5\,mm, and 2.0\,mm. In the ALMA images, the central blazar has been removed; The orange ellipse shows the FWHM of the synthesized beam and the white contours show the emission 3 $\sigma$ above the RMS noise. Except for the blazar, all the other sources are invisible in the optical image. }
   \label{fig:multiple_image}
\end{figure*}

\begin{table*}
  \centering
  \caption{Observed properties of the seven objects and the blazar.}
  \label{tab:continuum}
  \begin{tabular}{cccccccccc}
    \hline\hline
    ID & R.A. & Dec. & $z$ & $S_{\rm 2\,mm}$ & $S_{\rm 1.5\,mm}$ & $S_{\rm 1.2\,mm}$ & $S_{870\mu m}$ & Type\\ 
     & h:m:s & d:m:s & & (mJy) & (mJy) & (mJy) &  (mJy)\\ 
    \hline
HLW-1 & 02:17:02.73 & $-$08:21:02.1 & 2.240 & 0.20$\pm$0.05 & 0.42$\pm$0.20 & 1.44$\pm$0.05 & 4.00$\pm$0.19 & SMG\\
HLW-2 & 02:17:03.10 & $-$08:20:56.2 & 2.460 & 0.17$\pm$0.06 & 0.58$\pm$0.16 & 1.43$\pm$0.05 & 3.93$\pm$0.17 & SMG\\
HLW-3 & 02:17:02.94 & $-$08:20:39.2 & 1.238 & 0.30$\pm$0.08 & -- & 1.23$\pm$0.07 & 3.05$\pm$0.26 & SMG\\
HLW-4 & 02:17:02.47 & $-$08:20:56.9 & ? & -- & -- & 0.13$\pm$0.02 & 0.47$\pm$0.07 & DSFG\\
HLW-5 & 02:17:02.93 & $-$08:20:56.2 & 2.460? & -- & -- & 0.11$\pm$0.03 & 0.42$\pm$0.21 & DSFG\\
HLW-6 & 02:17:02.12 & $-$08:20:44.7 & 1.477 & -- & -- & 0.13$\pm$0.10 & 0.36$\pm$0.16 & DSFG\\
HLW-7 & 02:17:02.73 & $-$08:20:49.1 & ? & 0.98$\pm$0.05 & 0.79$\pm$0.10 & 0.53$\pm$0.08 & 0.43$\pm$0.11 & Synchrotron\\
\hline
J0217-0820 & 02:17:02.66 & $-$08:20:52.35 & 0.6065 & 281$\pm$14 & 226$\pm$10 & 157$\pm$8 & 115$\pm$6 & Blazar \\
    \hline
  \end{tabular}
\end{table*}

\section{ALMA data and their analysis}
\label{sec:data}

ALMACAL exploits `free' calibration data to monitor the ALMA calibration sources (most of which are blazars) and 
survey their immediate vicinities \citep{Zwaan2022}. 
The ALMA calibrators are distributed fairly randomly
across the sky, serving to calibrate the bandpass response, 
complex gains, the flux density scale, and the degree and 
angle of polarisation.
Overall, $\approx 20$\% of the available ALMA telescope
time is and has been spent on calibrators. 
For some calibrators, the depth reached by combining all the available visits is comparable with the deepest cosmological fields.
Since calibration scans typically cover many frequencies, 
the resulting spectral coverage can be considerable, 
offering the opportunity to search for spectral lines. 
Towards the calibrator, J1058+0133, for example, 
\citet{Oteo2017} found two SMGs and detected a host of spectral lines, 
yielding a remarkable spectral line energy distribution. Towards the calibrator J0238+1636, \citet{Klitsch2019a} observed multiple CO emission lines around one Ly\,$\alpha$ absorber, which shows excited interstellar medium.

\citet{Chen2023} conducted a multi-band
survey for dusty starbursts as part of the ALMACAL project. \JJ{} was
the densest region found among the available ALMACAL footprints. Until
August 2022, around 8.6\,h of data had accumulated. Combining
these data, ALMA has sampled the spectrum from band 3 to band 7.
In band 7, the 1-$\sigma$ depth reaches 13\,$\mu$Jy\,beam$^{-1}$.
Within $d=30''$ around the centre, seven continuum sources have been found ($>5\sigma$ significance),
including three at $S_{\rm 870\mu m}>1$\,mJy, which we will call traditional SMGs, following \citet{Hodge2020}, and
three fainter sources, which we refer to as DSFGs (see also Table \ref{tab:continuum}).

The data retrieval and calibration in \JJ{} is similar to the scheme described in
\citet{Oteo2016}. In summary,  we first collected
all the observations that used \JJ{} as a calibrator, then ran
the standard calibration offered by {\tt ScriptForPI.py} in each
project, splitting the calibrated data for \JJ{}. Within each
project, the flux density was calibrated by the dedicated
calibrator or scaled to the most recent online catalogue. 
Next, we self-calibrated the data in phase-only and in phase-and-amplitude mode
to improve the fidelity of the image.

Blazars tend to be variable at submm wavelengths
\citep[e.g.][]{Robson1983}. To minimise
the resulting effects, we used Common Astronomy Software Applications (CASA) tool, {\sc uvmodelfit}, to subtract
the central calibrator from each independent observation before imaging, 
adopting a point source model.

After these steps, we combined the observations to create the final
continuum image and datacube in each band, as follows.
We used two imaging cycles to create the continuum images. 
We first made dirty images of all the observations 
using {\sc tclean} in `mfs' mode, with zero clean iterations. 
Next, we rejected images with calibration issues and unsuccessful 
point-source subtraction, based on our visual inspection. 
Less than 2\% of the total data were lost. 
After that, we used the CASA task, {\sc statwt}, 
to re-calculate the weights, based on the noise in each observation. 
Finally, {\sc tclean} constructed the combined image,
wherein the {\sc auto-multithresh} algorithm searched for robust
emission during each major clean cycle.
We also tapered the image to 0.4 and 0.8\,arcsec spatial 
resolution (FWHM) to aid in the recovery of any extended sources. 
Our images of J0217$-$0820 are shown in Fig.~\ref{fig:multiple_image}.

We classified the continuum sources based on their multi-band colours (see Appendix\,\ref{appendix:classification}). 
Among the seven point-like sources, six have thermal spectra consistent with
dusty starbursts, and one is a synchrotron source. Continuum images
and measurements at different wavelengths are presented in
Fig.\,\ref{fig:multiple_image} and Table\,\ref{tab:continuum}. We name
the seven sources as the Calabash brothers, or Huluwa, 
abbreviated to HLW\footnote{There are seven Huluwa, 
each with different characteristics and abilities, see: 
\href{https://en.wikipedia.org/wiki/Calabash\_Brothers}{Calabash Brothers}}. There is also a
radio jet associated with the central blazar, heading south-east.

\begin{figure*}
   \centering
   \resizebox{0.8\hsize}{!}{\includegraphics{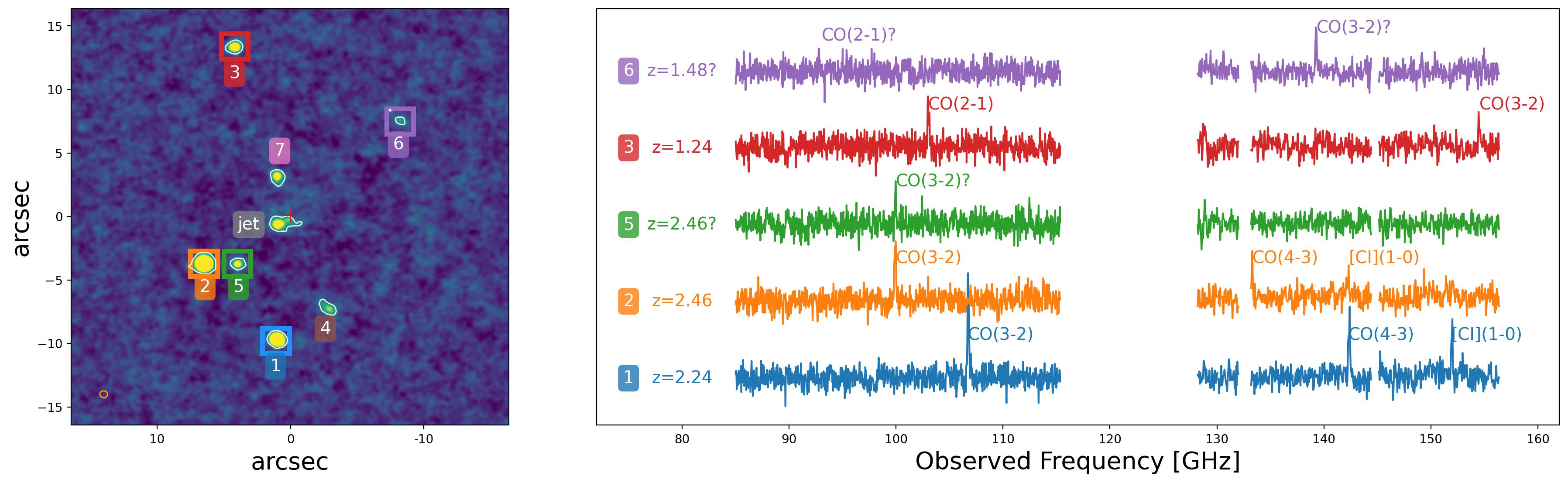}}
   \caption{Redshifts of the continuum sources in the \JJ{} field. Left:
     Continuum image at 870\,$\mu$m. Right: Confirmed spectral lines
     associated with those continuum sources in ALMA band 3 and band 4.
     A zoom version with the additional lines is available in Appendix Fig.~\ref{fig:spectra_zoom}.
     We have confirmed five redshift groups. 
     The central blazar lies at $z=0.6$,
     while the surrounding dusty starbursts are located at four
     different redshifts.}
   \label{fig:spectra}
\end{figure*}

To create the datacubes, we also followed two cycles. First, we used all the
available data to create a dirty datacube in each ALMA band. After
that, we extracted spectra at the positions of the seven continuum
sources. Since the calibrator always shares the same instrumental
configuration as the intended science target, the sensitivity can be
quite different from observation to observation, resulting in quite
different sensitivities at different frequencies. To make the best use
of the spectral data, we normalised the extracted spectra by the
sensitivities in each channel to create a `signal-to-noise (S/N)
spectrum' for each continuum source. These S/N spectra preserve
strong spectral lines whilst suppressing large spikes caused by the
low sensitivity at some frequencies. The resulting spectra are shown
in Fig.\,\ref{fig:spectra}.

Starting with the 5\,$\sigma$ peaks in our S/N spectra, we searched for
additional lines, assuming the original peak to be $^{12}$CO or
[C\,{\sc i}]. After searching for all the possible emission lines, we
returned to the original calibrated data and created a 1-GHz-wide
datacube for each line. Similarly to the cleaning process for our
continuum images, {\sc auto-multishresh} was applied to identify
channels with robust emission. We stopped cleaning when the peak
emission fell below the 2$\sigma$ noise level of the residual image.

\section{Over-densities}
\label{sec:overdensity}

We define the over-density of galaxies as follows:

\begin{equation}
    \delta_{\rm gal} = \frac{n_{\rm gal}- \langle n_{\rm gal}\rangle}{\langle n_{\rm gal}\rangle},
\end{equation}

\noindent
where $\delta_{\rm gal}$ is the galaxy over-density, $n_{\rm gal}$ is
the number of galaxies found in the field of interest and
$\langle n_{\rm gal} \rangle$ is the average number of galaxies from
the adopted reference survey.

Towards \JJ{}, we detected six DSFGs with
$S_{\rm 870\mu m}>0.3$\,mJy, with three of them being classical SMGs with
$S_{\rm 870\mu m}>3$\,mJy. Based on the number counts from large
surveys at 870\,$\mu$m (\citealt{Chen2023}, see also
\citealt{Stach2018, Bethermin2020, Simpson2020}), the expected average
number of DSFGs in the FoV of ALMA band 7 ($d=30''$) is 
${\langle n_{\rm gal}\rangle}\approx0.050\pm0.002$ at $S_{\rm 870\mu m}>3$\,mJy and ${\langle n_{\rm gal}\rangle}\approx0.67\pm0.08$ at $S_{\rm 870\mu m}>0.3$\,mJy.
Then, the inferred over-density is $\delta_{\rm gal}\approx$59 and $\delta_{\rm gal}\approx$8.0, respectively.
If we were to count the synchrotron source, the over-density would be even higher.
Empirically, finding $\delta_{\rm gal}\gtrsim 8$ indicates strongly that
a field is not consistent with the reference field, and likely
contains a proto-cluster \citep[e.g.][]{Chiang2013, Lovell2018}.
Solely based on its over-density, \JJ{} is thus an extremely promising
proto-cluster candidate, similar to DRC \citep{Lewis2018}.

\section{Redshift constraints}
\label{sec:redshifts}

We identified 15 spectral lines in this field thanks to the wide
spectral coverage of ALMACAL. We list all the confirmed spectral lines
and their velocity-integrated flux densities in
Appendix Table\,\ref{tab:spectral_lines}.

The three brightest  SMGs (HLW-1,
HLW-2, HLW-3) have at least three spectral lines each, sufficient to
tie down their redshifts unambiguously. HLW-4 displays no robust
spectral lines. HLW-5 has only one robust spectral line, but it is at
the same frequency as a line seen from its neighbour, HLW-2, so they are
likely an interacting pair, as is common or possibly even ubiquitous
amongst SMGs \citep{Engel2010}. HLW-6 has one strong spectral line
($>\!5\sigma$) as well as a weak line at 92.8\,GHz, consistent with
CO(2--1) and (3--2) at z=1.477. Even without the second,
weaker line, it is clear that HLW-6 is not associated with any other
galaxies in this field.

Considering its synchrotron spectrum, HLW-7 is likely at $z=0.6065$ 
\citep[as determined for the central blazar by 
the Sloan Digital Sky Survey --][]{Albareti2017}, 
that is,\ it may well be another jet hotspot or it can be a companion to the blazar, as they are close enough together to have possibly triggered each
other's AGN activity.

\section{Discussion and conclusions}
\label{sec:discussion}

\begin{figure}
   \centering
   \resizebox{0.8\hsize}{!}{\includegraphics{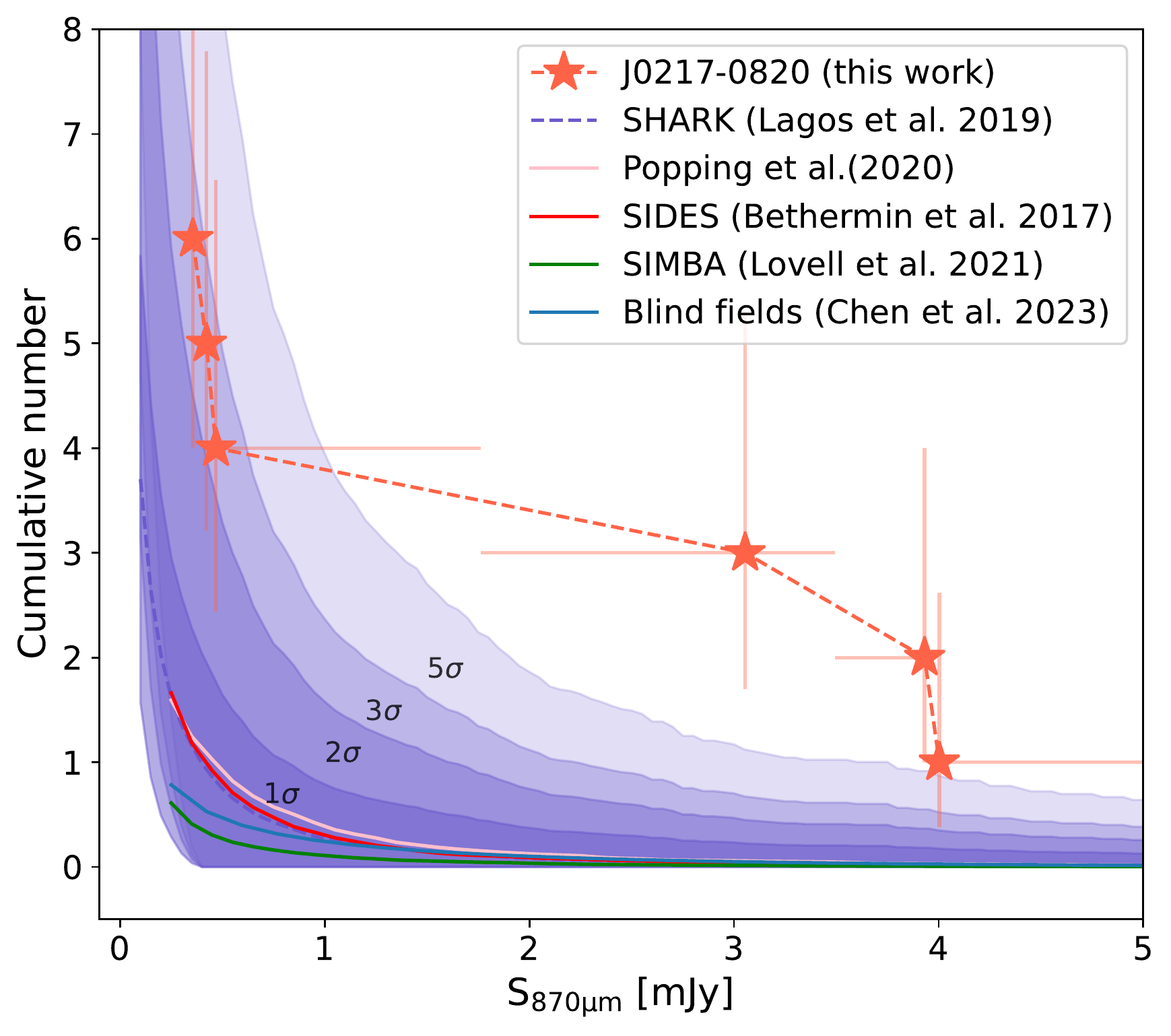}}
   \caption{Projected number of galaxies in various simulations. We search for galaxies following the same configuration of ALMA band 7 ($d=30''$).
     We include the results from the lightcones of semi-empirical models \citep{Bethermin2017, Popping2020}, semi-analytic models \citep{Lagos2019,Lagos2020}, and hydrodynamical simulations \citep{Lovell2021}. 
     The shadows show the 1\,$\sigma$, 2\,$\sigma$, 3\,$\sigma$, and 5\,$\sigma$ scatter in the lightcones of the SHARK simulation \citep{Lagos2019}.
     Error bars indicate the Poisson errors.
     Within the simulations it is very rare ($>5\sigma$) to find the number of SMGs seen around \JJ{}, but the deeper the observation goes, the stronger the projection effect.}
   \label{fig:cosmology}
\end{figure}

\JJ{} was discovered within the 45-arcmin$^2$ ALMACAL band 7 footprint, which is about 100\,Mpc$^2$ (comoving coordinates) at $z\sim2$.
It is several orders of magnitude smaller than the parent surveys that led to the discoveries of the two extremes of confirmed proto-cluster cores \citep{Oteo2018, Miller2018}, because we started from data with a better spatial resolution by an order of magnitude.
The negative $K$ correction has greatly benefited the search for DSFGs across a wide redshift range, but because of this it also amplifies the projection effects \citep{Hayward2013b,Lovell2021}. 
Clearly, given what we have found here, over-densities such as that around \JJ{} could contaminate the search for genuine proto-cluster cores in future deep submm/mm sky surveys.

Strong lensing could amplify the flux density of the background sources and boost the number counts in the vicinity.
Around the foreground blazar, we did not find other galaxies at the same redshift or share similar colours in the Dark Energy Spectroscopic Instrument (DESI) and VISTA/VIRCam images.
It is therefore unlikely that this over-density is caused by the strong lensing of a foreground galaxy cluster.
Meanwhile, the morphologies of the submm sources do not support the idea that they have been strongly lensed by the blazar.

We considered whether it was possible to find such alignments in cosmological simulations.
To address the question, we made mock light cones based on recent simulations that
predict submm/mm emission characteristics, then we searched those light cones for
projected over-densities. We adopted the simulations from
semi-analytical models \citep[SAMs --][]{Lagos2019, Lagos2020},
semi-empirical models \citep{Bethermin2017, Popping2020}, and
hydro-dynamical simulations \citep{Lovell2021}.  
The results are presented in Fig.\,\ref{fig:cosmology} and we offer more details in
Appendix \ref{appendix:simulations}.

We first note that almost all of the simulations predict much lower
numbers of traditional, bright SMGs, with a significance higher than 5$\sigma$.
This inconsistency could indicate that what we have found here is a real, rare alignment.
However, the failure to reproduce the number density of SMGs has long been reported in the literature. 
Possible explanations are rooted in the stellar initial mass function \citep[e.g.][]{Baugh2005, Lacey2016}, 
stellar and AGN feedback \citep{Lovell2021, Hayward2021} as well as various dust-related
models \citep[including production, destruction and temperature; 
see a detailed discussion in][]{Safarzadeh2017, Lagos2019, Lovell2021}.
Therefore, if such a chance alignment is much higher than what has been predicted,
this discrepancy may also suggest that we could
have under-predicted the cosmic variance of submm surveys.

At the fainter end, the probability of finding a similar overdensity is much higher than at the bright end.
At $S_{870\mu m}\sim0.1$mJy, the chance of finding more than six DSFGs is close to 5\%.
Considering the Poisson process on its own, the probability is only around 0.1\%,
which indicates that clustered structures in the Universe contribute to the projection effects.
This effect should be strongest at submm/mm wavelengths, due to the large redshift space probed in the submm/mm windows.
This cautions against the use of a simplistic statistical 
approach to select proto-clusters and quantify cosmic variance in deep submm surveys.

In summary, although \JJ{} is an extremely over-dense region -- where its over-density, $\delta_{\rm gal}\simeq 8$, 
is comparable to that of known extreme proto-cluster cores -- we have determined that this is almost entirely due to chance alignments. 
In future deep galaxy surveys, for instance, those with ALMA in the submm/mm bands, similar
projection effects may contaminate the search for proto-clusters and contribute to the cosmic variance.

\begin{acknowledgements}
\label{acknowledgements}
We are grateful for valuable feedback from an anonymous referee and inspiring discussions with Ian Smail, Gerg\"o Popping and Roland Szakacs, all of which improved the manuscript.
This work is funded by the Deutsche Forschungsgemeinschaft (DFG, German Research Foundation) under Germany's Excellence Strategy -- EXC-2094 -- 390783311.  This paper makes use of the ALMA data (see the full list in Appendix:\,\ref{appendix:almaid}). ALMA is a partnership of ESO (representing its member states), NSF (USA) and NINS (Japan), together with NRC (Canada), MOST and ASIAA (Taiwan), and KASI (Republic of Korea), in cooperation with the Republic of Chile. The Joint ALMA Observatory is operated by ESO, AUI/NRAO and NAOJ.

All the data used in the work is publicly available from the ALMA science archive \hyperlink{ALMA Science Archive}{https://almascience.eso.org/alma-data}.
\end{acknowledgements}

\bibliographystyle{aa} 
\bibliography{cautionary_tale_protocluster2} 

\begin{thebibliography}{77}
\expandafter\ifx\csname natexlab\endcsname\relax\def\natexlab#1{#1}\fi

\bibitem[{Albareti {et~al.}(2017)Albareti, Allende~Prieto, Almeida, Anders,
  Anderson, Andrews, {Arag{\'o}n-Salamanca}, {Argudo-Fern{\'a}ndez}, Armengaud,
  Aubourg, {Avila-Reese}, Badenes, Bailey, Barbuy, Barger,
  {Barrera-Ballesteros}, Bartosz, Basu, Bates, Battaglia, Baumgarten, Baur,
  Bautista, Beers, Belfiore, Bershady, {Bertran de Lis}, Bird, Bizyaev, Blanc,
  Blanton, Blomqvist, Bolton, Borissova, Bovy, Brandt, Brinkmann, Brownstein,
  Bundy, Burtin, Busca, Camacho~Chavez, Cano~D{\'i}az, Cappellari, Carrera,
  Chen, Cherinka, Cheung, Chiappini, Chojnowski, Chuang, Chung, Cirolini,
  Clerc, Cohen, Comerford, Comparat, {Correa do Nascimento}, Cousinou, Covey,
  Crane, Croft, Cunha, Darling, Davidson, Dawson, Da~Costa, Da~Silva~Ilha,
  Deconto~Machado, Delubac, De~Lee, {De la Macorra}, {De la Torre},
  {Diamond-Stanic}, Donor, Downes, Drory, Du, {Du Mas des Bourboux}, Dwelly,
  Ebelke, Eigenbrot, Eisenstein, Elsworth, Emsellem, Eracleous, Escoffier,
  Evans, {Falc{\'o}n-Barroso}, Fan, Favole, {Fernandez-Alvar},
  {Fernandez-Trincado}, Feuillet, Fleming, {Font-Ribera}, Freischlad,
  Frinchaboy, Fu, Gao, Garcia, {Garcia-Dias}, {Garcia-Hern{\'a}ndez},
  Garcia~P{\'e}rez, Gaulme, Ge, Geisler, Gillespie, Gil~Marin, Girardi,
  Goddard, Gomez Maqueo~Chew, {Gonzalez-Perez}, Grabowski, Green, Grier, Grier,
  Guo, Guy, Hagen, Hall, Harding, Harley, Hasselquist, Hawley, Hayes, Hearty,
  Hekker, Hernandez~Toledo, Ho, Hogg, {Holley-Bockelmann}, Holtzman, Holzer,
  Hu, Huber, Hutchinson, Hwang, {Ibarra-Medel}, Ivans, Ivory, Jaehnig, Jensen,
  Johnson, Jones, Jullo, Kallinger, Kinemuchi, Kirkby, Klaene, Kneib,
  Kollmeier, Lacerna, Lane, Lang, Laurent, Law, Leauthaud, Le~Goff, Li, Li, Li,
  Li, Liang, Liang, Lima, Lin, Lin, Lin, Liu, Long, Lucatello, MacDonald,
  MacLeod, Mackereth, Mahadevan, Maia, Maiolino, Majewski, Malanushenko,
  Malanushenko, Mallmann, Manchado, Maraston, {Marques-Chaves},
  Martinez~Valpuesta, Masters, Mathur, McGreer, Merloni, Merrifield,
  M{\'e}sz{\'a}ros, Meza, Miglio, Minchev, Molaverdikhani, {Montero-Dorta},
  Mosser, Muna, Myers, Nair, Nandra, Ness, Newman, Nichol, Nidever, Nitschelm,
  O'Connell, Oravetz, Oravetz, Pace, Padilla, {Palanque-Delabrouille}, Pan,
  Parejko, Paris, Park, Peacock, Peirani, {Pellejero-Ibanez}, Penny, Percival,
  Percival, {Perez-Fournon}, Petitjean, Pieri, Pinsonneault, Pisani, Prada,
  Prakash, {Price-Jones}, Raddick, Rahman, Raichoor, Barboza~Rembold, Reyna,
  Rich, Richstein, Ridl, Riffel, Riffel, Rix, Robin, Rockosi,
  {Rodr{\'i}guez-Torres}, Rodrigues, Roe, Roman~Lopes,
  {Rom{\'a}n-Z{\'u}{\~n}iga}, Ross, Rossi, Ruan, Ruggeri, Runnoe,
  {Salazar-Albornoz}, Salvato, Sanchez, Sanchez, {Sanchez-Gallego}, Santiago,
  Schiavon, Schimoia, Schlafly, Schlegel, Schneider, Sch{\"o}nrich, Schultheis,
  Schwope, Seo, Serenelli, Sesar, Shao, Shetrone, Shull, Silva~Aguirre,
  Skrutskie, Slosar, Smith, Smith, Sobeck, Somers, Souto, Stark, Stassun,
  Steinmetz, Stello, Storchi~Bergmann, Strauss, Streblyanska, Stringfellow,
  Suarez, Sun, {Taghizadeh-Popp}, Tang, Tao, Tayar, Tembe, Thomas, Tinker,
  Tojeiro, Tremonti, Troup, Trump, {Unda-Sanzana}, Valenzuela, {Van den Bosch},
  {Vargas-Maga{\~n}a}, Vazquez, Villanova, Vivek, Vogt, Wake, Walterbos, Wang,
  Wang, Weaver, Weijmans, Weinberg, Westfall, Whelan, Wilcots, Wild, Williams,
  Wilson, {Wood-Vasey}, Wylezalek, Xiao, Yan, Yang, Ybarra, Yeche, Yuan,
  Zakamska, Zamora, Zasowski, Zhang, Zhao, Zhao, Zheng, Zheng, Zhou, Zhu, Zinn,
  \& Zou}]{Albareti2017}
Albareti, F.~D., Allende~Prieto, C., Almeida, A., {et~al.} 2017, ApJ, 233, 25

\bibitem[{Alberts \& Noble(2022)}]{Alberts2022}
Alberts, S. \& Noble, A. 2022, Universe, 8, 554

\bibitem[{Barger {et~al.}(1998)Barger, Cowie, Sanders, Fulton, Taniguchi, Sato,
  Kawara, \& Okuda}]{Barger1998}
Barger, A.~J., Cowie, L.~L., Sanders, D.~B., {et~al.} 1998, Nature, 394, 248

\bibitem[{Baugh {et~al.}(2005)Baugh, Lacey, Frenk, Granato, Silva, Bressan,
  Benson, \& Cole}]{Baugh2005}
Baugh, C.~M., Lacey, C.~G., Frenk, C.~S., {et~al.} 2005, MNRAS, 356, 1191

\bibitem[{Behroozi {et~al.}(2019)Behroozi, Wechsler, Hearin, \&
  Conroy}]{Behroozi2019}
Behroozi, P., Wechsler, R.~H., Hearin, A.~P., \& Conroy, C. 2019, MNRAS, 488,
  3143

\bibitem[{B{\'e}thermin {et~al.}(2020)B{\'e}thermin, Fudamoto, Ginolfi,
  Loiacono, Khusanova, Capak, Cassata, Faisst, Le~F{\`e}vre, Schaerer,
  Silverman, Yan, Amorin, Bardelli, Boquien, Cimatti, Davidzon,
  {Dessauges-Zavadsky}, Fujimoto, Gruppioni, Hathi, Ibar, Jones, Koekemoer,
  Lagache, Lemaux, Moreau, Oesch, Pozzi, Riechers, Talia, Toft, Vallini,
  Vergani, Zamorani, \& Zucca}]{Bethermin2020}
B{\'e}thermin, M., Fudamoto, Y., Ginolfi, M., {et~al.} 2020, A\&A, 643, A2

\bibitem[{B{\'e}thermin {et~al.}(2017)B{\'e}thermin, Wu, Lagache, Davidzon,
  Ponthieu, Cousin, Wang, Dor{\'e}, Daddi, \& Lapi}]{Bethermin2017}
B{\'e}thermin, M., Wu, H.-Y., Lagache, G., {et~al.} 2017, A\&A, 607, A89

\bibitem[{Bond {et~al.}(1996)Bond, Kofman, \& Pogosyan}]{Bond1996}
Bond, J.~R., Kofman, L., \& Pogosyan, D. 1996, Nature, 380, 603

\bibitem[{Calvi {et~al.}(2023)Calvi, Castignani, \& Dannerbauer}]{Calvi2023}
Calvi, R., Castignani, G., \& Dannerbauer, H. 2023, Submillimeter Galaxies do
  trace Galaxy Protoclusters

\bibitem[{Capak {et~al.}(2011)Capak, Riechers, Scoville, Carilli, Cox, Neri,
  Robertson, Salvato, Schinnerer, Yan, Wilson, Yun, Civano, Elvis, Karim,
  Mobasher, \& Staguhn}]{Capak2011}
Capak, P.~L., Riechers, D., Scoville, N.~Z., {et~al.} 2011, Nature, 470, 233

\bibitem[{Chapman {et~al.}(2015)Chapman, Bertoldi, Smail, Steidel, Blain,
  Geach, Gurwell, Ivison, Petitpas, \& Reddy}]{Chapman2015}
Chapman, S.~C., Bertoldi, F., Smail, I., {et~al.} 2015, MNRAS, 453, 951

\bibitem[{Chen {et~al.}(2023)Chen, Ivison, Zwaan, Smail, Klitsch, P{\'e}roux,
  Popping, Biggs, Szakacs, Hamanowicz, \& Lagos}]{Chen2023}
Chen, J., Ivison, R.~J., Zwaan, M.~A., {et~al.} 2023, MNRAS, 518, 1378

\bibitem[{Chiang {et~al.}(2013)Chiang, Overzier, \& Gebhardt}]{Chiang2013}
Chiang, Y.-K., Overzier, R., \& Gebhardt, K. 2013, ApJ, 779, 127

\bibitem[{Dannerbauer {et~al.}(2014)Dannerbauer, Kurk, De~Breuck, Wylezalek,
  Santos, Koyama, Seymour, Tanaka, Hatch, Altieri, Coia, Galametz, Kodama,
  Miley, R{\"o}ttgering, {Sanchez-Portal}, Valtchanov, Venemans, \&
  Ziegler}]{Dannerbauer2014}
Dannerbauer, H., Kurk, J.~D., De~Breuck, C., {et~al.} 2014, A\&A, 570, A55

\bibitem[{Dav{\'e} {et~al.}(2019)Dav{\'e}, {Angl{\'e}s-Alc{\'a}zar}, Narayanan,
  Li, Rafieferantsoa, \& Appleby}]{Dave2019a}
Dav{\'e}, R., {Angl{\'e}s-Alc{\'a}zar}, D., Narayanan, D., {et~al.} 2019,
  MNRAS, 486, 2827

\bibitem[{Dekel {et~al.}(2009)Dekel, Sari, \& Ceverino}]{Dekel2009}
Dekel, A., Sari, R., \& Ceverino, D. 2009, AJ, 703, 785

\bibitem[{Engel {et~al.}(2010)Engel, Tacconi, Davies, Neri, Smail, Chapman,
  Genzel, Cox, Greve, Ivison, Blain, Bertoldi, \& Omont}]{Engel2010}
Engel, H., Tacconi, L.~J., Davies, R.~I., {et~al.} 2010, ApJ, 724, 233

\bibitem[{Hatch {et~al.}(2011)Hatch, Kurk, Pentericci, Venemans, Kuiper, Miley,
  \& R{\"o}ttgering}]{Hatch2011}
Hatch, N.~A., Kurk, J.~D., Pentericci, L., {et~al.} 2011, MNRAS, 415, 2993

\bibitem[{Hayashi {et~al.}(2012)Hayashi, Kodama, Tadaki, Koyama, \&
  Tanaka}]{Hayashi2012}
Hayashi, M., Kodama, T., Tadaki, K.-i., Koyama, Y., \& Tanaka, I. 2012, ApJ,
  757, 15

\bibitem[{Hayward {et~al.}(2013)Hayward, Behroozi, Somerville, Primack, Moreno,
  \& Wechsler}]{Hayward2013b}
Hayward, C.~C., Behroozi, P.~S., Somerville, R.~S., {et~al.} 2013, MNRAS, 434,
  2572

\bibitem[{Hayward {et~al.}(2021)Hayward, Sparre, Chapman, Hernquist, Nelson,
  Pakmor, Pillepich, Springel, Torrey, Vogelsberger, \&
  Weinberger}]{Hayward2021}
Hayward, C.~C., Sparre, M., Chapman, S.~C., {et~al.} 2021, MNRAS, 502, 2922

\bibitem[{Helton {et~al.}(2023)Helton, Sun, Woodrum, Hainline, Willmer, Rieke,
  Rieke, Tacchella, Robertson, Johnson, Alberts, Eisenstein, Hausen,
  Bonaventura, Bunker, Charlot, Curti, {Curtis-Lake}, Looser, Maiolino,
  Willott, Witstok, Boyett, Chen, Egami, Endsley, Hviding, Jaffe, Ji, Lyu, \&
  Sandles}]{Helton2023}
Helton, J.~M., Sun, F., Woodrum, C., {et~al.} 2023, The JWST Advanced Deep
  Extragalactic Survey: Discovery of an Extreme Galaxy Overdensity at \$z =
  5.4\$ with JWST/NIRCam in GOODS-S

\bibitem[{Hodge \& {da Cunha}(2020)}]{Hodge2020}
Hodge, J.~A. \& {da Cunha}, E. 2020, R. Soc. Open Sci., 7, 200556

\bibitem[{Hughes {et~al.}(1998)Hughes, Serjeant, Dunlop, {Rowan-Robinson},
  Blain, Mann, Ivison, Peacock, Efstathiou, Gear, Oliver, Lawrence, Longair,
  Goldschmidt, \& Jenness}]{Hughes1998}
Hughes, D.~H., Serjeant, S., Dunlop, J., {et~al.} 1998, Nature, 394, 241

\bibitem[{Ivison {et~al.}(2020)Ivison, Biggs, Bremer, Arumugam, \&
  Dunne}]{Ivison2020}
Ivison, R.~J., Biggs, A.~D., Bremer, M., Arumugam, V., \& Dunne, L. 2020,
  MNRAS, 496, 4358

\bibitem[{Ivison {et~al.}(2016)Ivison, Lewis, Weiss, Arumugam, Simpson,
  Holland, Maddox, Dunne, Valiante, {van der Werf}, Omont, Dannerbauer, Smail,
  Bertoldi, Bremer, Bussmann, Cai, Clements, Cooray, De~Zotti, Eales, Fuller,
  {Gonzalez-Nuevo}, Ibar, Negrello, Oteo, {P{\'e}rez-Fournon}, Riechers,
  Stevens, Swinbank, \& Wardlow}]{Ivison2016a}
Ivison, R.~J., Lewis, A. J.~R., Weiss, A., {et~al.} 2016, ApJ, 832, 78

\bibitem[{Jin {et~al.}(2023)Jin, Sillassen, Magdis, Vijayan, Brammer, Kokorev,
  Weaver, Gobat, {Gim{\'e}nez-Arteaga}, Valentino, Brinch,
  {G{\'o}mez-Guijarro}, Shuntov, Toft, Greve, \& Blanquez~Sese}]{Jin2023}
Jin, S., Sillassen, N.~B., Magdis, G.~E., {et~al.} 2023, A\&A, 670, L11

\bibitem[{Klitsch {et~al.}(2019)Klitsch, P{\'e}roux, Zwaan, Smail, Nelson,
  Popping, Chen, Diemer, Ivison, Allison, Muller, Swinbank, Hamanowicz, Biggs,
  \& Dutta}]{Klitsch2019a}
Klitsch, A., P{\'e}roux, C., Zwaan, M.~A., {et~al.} 2019, Monthly Notices of
  the Royal Astronomical Society, 490, 1220

\bibitem[{Kurk {et~al.}(2004)Kurk, Pentericci, Overzier, R{\"o}ttgering, \&
  Miley}]{Kurk2004a}
Kurk, J.~D., Pentericci, L., Overzier, R.~A., R{\"o}ttgering, H. J.~A., \&
  Miley, G.~K. 2004, A\&A, 428, 817

\bibitem[{Kurk {et~al.}(2000)Kurk, R{\"o}ttgering, Pentericci, Miley, {van
  Breugel}, Carilli, Ford, Heckman, McCarthy, \& Moorwood}]{Kurk2000}
Kurk, J.~D., R{\"o}ttgering, H. J.~A., Pentericci, L., {et~al.} 2000, A\&A,
  358, L1

\bibitem[{Lacey {et~al.}(2016)Lacey, Baugh, Frenk, Benson, Bower, Cole,
  {Gonzalez-Perez}, Helly, Lagos, \& Mitchell}]{Lacey2016}
Lacey, C.~G., Baugh, C.~M., Frenk, C.~S., {et~al.} 2016, MNRAS, 462, 3854

\bibitem[{Lagos {et~al.}(2020)Lagos, {da Cunha}, Robotham, Obreschkow,
  Valentino, Fujimoto, Magdis, \& Tobar}]{Lagos2020}
Lagos, C. d.~P., {da Cunha}, E., Robotham, A. S.~G., {et~al.} 2020, MNRAS, 499,
  1948

\bibitem[{Lagos {et~al.}(2019)Lagos, Robotham, Trayford, Tobar, Bravo,
  Bellstedt, Davies, Driver, Elahi, Obreschkow, \& Power}]{Lagos2019}
Lagos, C. d.~P., Robotham, A. S.~G., Trayford, J.~W., {et~al.} 2019, MNRAS,
  489, 4196

\bibitem[{Lammers {et~al.}(2022)Lammers, Hill, Lim, Scott, Ca{\~n}ameras, \&
  Dole}]{Lammers2022}
Lammers, C., Hill, R., Lim, S., {et~al.} 2022, MNRAS, 514, 5004

\bibitem[{Laporte {et~al.}(2022)Laporte, Zitrin, Dole, {Roberts-Borsani},
  Furtak, \& Witten}]{Laporte2022}
Laporte, N., Zitrin, A., Dole, H., {et~al.} 2022, A\&A, 667, L3

\bibitem[{Lewis {et~al.}(2018)Lewis, Ivison, Best, Simpson, Weiss, Oteo, Zhang,
  Arumugam, Bremer, Chapman, Clements, Dannerbauer, Dunne, Eales, Maddox,
  Oliver, Omont, Riechers, Serjeant, Valiante, Wardlow, {van der Werf}, \&
  De~Zotti}]{Lewis2018}
Lewis, A. J.~R., Ivison, R.~J., Best, P.~N., {et~al.} 2018, ApJ, 862, 96

\bibitem[{Li {et~al.}(2023)Li, Wang, Fan, Wu, Jiang, Ba{\~n}ados, Venemans,
  Shao, Li, Wagg, Decarli, Mazzucchelli, Omont, Bertoldi, Johnson, \&
  Conselice}]{Li2023}
Li, Q., Wang, R., Fan, X., {et~al.} 2023, (SHERRY) JCMT-SCUBA2 High Redshift
  Bright Quasar Survey -- II: the environment of z\textasciitilde 6 quasars in
  sub-millimeter band

\bibitem[{Li {et~al.}(2020)Li, Wang, Fan, Wu, Jiang, Ba{\~n}ados, Venemans,
  Shao, Li, Zhang, Zhang, Wagg, Decarli, Mazzucchelli, Omont, \&
  Bertoldi}]{Li2020}
Li, Q., Wang, R., Fan, X., {et~al.} 2020, ApJ, 900, 12

\bibitem[{Lilly {et~al.}(1999)Lilly, Eales, Gear, Hammer, Le~F{\`e}vre,
  Crampton, Bond, \& Dunne}]{Lilly1999}
Lilly, S.~J., Eales, S.~A., Gear, W. K.~P., {et~al.} 1999, ApJ, 518, 641

\bibitem[{Lovell {et~al.}(2021)Lovell, Geach, Dav{\'e}, Narayanan, \&
  Li}]{Lovell2021}
Lovell, C.~C., Geach, J.~E., Dav{\'e}, R., Narayanan, D., \& Li, Q. 2021,
  MNRAS, 502, 772

\bibitem[{Lovell {et~al.}(2018)Lovell, Thomas, \& Wilkins}]{Lovell2018}
Lovell, C.~C., Thomas, P.~A., \& Wilkins, S.~M. 2018, MNRAS, 474, 4612

\bibitem[{MacKenzie {et~al.}(2017)MacKenzie, Scott, Bianconi, Clements, Dole,
  {Flores-Cacho}, Guery, Kneissl, Lagache, Marleau, Montier, Nesvadba,
  Pointecouteau, \& Soucail}]{MacKenzie2017}
MacKenzie, T.~P., Scott, D., Bianconi, M., {et~al.} 2017, MNRAS, 468, 4006

\bibitem[{Matsuda {et~al.}(2004)Matsuda, Yamada, Hayashino, Tamura, Yamauchi,
  Ajiki, Fujita, Murayama, Nagao, Ohta, Okamura, Ouchi, Shimasaku, Shioya, \&
  Taniguchi}]{Matsuda2004}
Matsuda, Y., Yamada, T., Hayashino, T., {et~al.} 2004, AJ, 128, 569

\bibitem[{Meyer {et~al.}(2022)Meyer, Decarli, Walter, Li, Wang, Mazzucchelli,
  Ba{\~n}ados, Farina, \& Venemans}]{Meyer2022}
Meyer, R.~A., Decarli, R., Walter, F., {et~al.} 2022, ApJ, 927, 141

\bibitem[{Miley \& De~Breuck(2008)}]{Miley2008}
Miley, G. \& De~Breuck, C. 2008, Astron Astrophys Rev, 15, 67

\bibitem[{Miller {et~al.}(2018)Miller, Chapman, Aravena, Ashby, Hayward,
  Vieira, Wei{\ss}, Babul, B{\'e}thermin, Bradford, Brodwin, Carlstrom, Chen,
  Cunningham, De~Breuck, Gonzalez, Greve, Harnett, Hezaveh, Lacaille, Litke,
  Ma, Malkan, Marrone, Morningstar, Murphy, Narayanan, Pass, Perry, Phadke,
  Rennehan, Rotermund, Simpson, Spilker, Sreevani, Stark, Strandet, \&
  Strom}]{Miller2018}
Miller, T.~B., Chapman, S.~C., Aravena, M., {et~al.} 2018, Nature, 556, 469

\bibitem[{Morishita {et~al.}(2023)Morishita, {Roberts-Borsani}, Treu, Brammer,
  Mason, Trenti, Vulcani, Wang, Acebron, Bah{\'e}, Bergamini, Boyett, Bradac,
  Calabr{\`o}, Castellano, Chen, De~Lucia, Filippenko, Fontana, Glazebrook,
  Grillo, Henry, Jones, Kelly, Koekemoer, Leethochawalit, Lu, Marchesini,
  Mascia, Mercurio, Merlin, Metha, Nanayakkara, Nonino, Paris, Pentericci,
  Rosati, Santini, Strait, Vanzella, Windhorst, \& Xie}]{Morishita2023}
Morishita, T., {Roberts-Borsani}, G., Treu, T., {et~al.} 2023, ApJ, 947, L24

\bibitem[{Nowotka {et~al.}(2022)Nowotka, Chen, Battaia, Fumagalli, Cai, Lusso,
  Prochaska, \& Yang}]{Nowotka2022}
Nowotka, M., Chen, C.-C., Battaia, F.~A., {et~al.} 2022, A\&A, 658, A77

\bibitem[{Oteo {et~al.}(2018)Oteo, Ivison, Dunne, {Manilla-Robles}, Maddox,
  Lewis, de~Zotti, Bremer, Clements, Cooray, Dannerbauer, Eales, Greenslade,
  Omont, {Perez{\textendash}Fourn{\'o}n}, Riechers, Scott, {van der Werf},
  Weiss, \& Zhang}]{Oteo2018}
Oteo, I., Ivison, R.~J., Dunne, L., {et~al.} 2018, ApJ, 856, 72

\bibitem[{Oteo {et~al.}(2016)Oteo, Zwaan, Ivison, Smail, \& Biggs}]{Oteo2016}
Oteo, I., Zwaan, M.~A., Ivison, R.~J., Smail, I., \& Biggs, A.~D. 2016, ApJ,
  822, 36

\bibitem[{Oteo {et~al.}(2017)Oteo, Zwaan, Ivison, Smail, \& Biggs}]{Oteo2017}
Oteo, I., Zwaan, M.~A., Ivison, R.~J., Smail, I., \& Biggs, A.~D. 2017, ApJ,
  837, 182

\bibitem[{Ouchi {et~al.}(2005)Ouchi, Shimasaku, Akiyama, Sekiguchi, Furusawa,
  Okamura, Kashikawa, Iye, Kodama, Saito, Sasaki, Simpson, Takata, Yamada,
  Yamanoi, Yoshida, \& Yoshida}]{Ouchi2005}
Ouchi, M., Shimasaku, K., Akiyama, M., {et~al.} 2005, ApJ, 620, L1

\bibitem[{Overzier(2016)}]{Overzier2016}
Overzier, R.~A. 2016, A\&ARv, 24, 14

\bibitem[{Overzier {et~al.}(2013)Overzier, Nesvadba, Dijkstra, Hatch, Lehnert,
  {Villar-Mart{\'i}n}, Wilman, \& Zirm}]{Overzier2013}
Overzier, R.~A., Nesvadba, N. P.~H., Dijkstra, M., {et~al.} 2013, ApJ, 771, 89

\bibitem[{Pentericci {et~al.}(2002)Pentericci, Kurk, Carilli, Harris, Miley, \&
  R{\"o}ttgering}]{Pentericci2002}
Pentericci, L., Kurk, J.~D., Carilli, C.~L., {et~al.} 2002, A\&A, 396, 109

\bibitem[{Popping {et~al.}(2020)Popping, Walter, Behroozi,
  {Gonz{\'a}lez-L{\'o}pez}, Hayward, Somerville, {van der Werf}, Aravena,
  Assef, Boogaard, Bauer, Cortes, Cox, {D{\'i}az-Santos}, Decarli, Franco,
  Ivison, Riechers, Rix, \& Weiss}]{Popping2020}
Popping, G., Walter, F., Behroozi, P., {et~al.} 2020, ApJ, 891, 135

\bibitem[{Robson {et~al.}(1983)Robson, Gear, Clegg, Ade, Smith, Griffin, Nolt,
  Radostitz, \& Howard}]{Robson1983}
Robson, E.~I., Gear, W.~K., Clegg, P.~E., {et~al.} 1983, Nature, 305, 194

\bibitem[{Robson {et~al.}(2014)Robson, Ivison, Smail, Holland, Geach, Gibb,
  Riechers, Ade, Bintley, Bock, Chapin, Chapman, Clements, Conley, Cooray,
  Dunlop, Farrah, Fich, Fu, Jenness, Laporte, Oliver, Omont,
  {P{\'e}rez-Fournon}, Scott, Swinbank, \& Wardlow}]{Robson2014}
Robson, E.~I., Ivison, R.~J., Smail, I., {et~al.} 2014, ApJ, 793, 11

\bibitem[{Safarzadeh {et~al.}(2017)Safarzadeh, Lu, \& Hayward}]{Safarzadeh2017}
Safarzadeh, M., Lu, Y., \& Hayward, C.~C. 2017, MNRAS, 472, 2462

\bibitem[{Seymour {et~al.}(2007)Seymour, Stern, De~Breuck, Vernet, Rettura,
  Dickinson, Dey, Eisenhardt, Fosbury, Lacy, McCarthy, Miley,
  {Rocca-Volmerange}, R{\"o}ttgering, Stanford, Teplitz, {van Breugel}, \&
  Zirm}]{Seymour2007}
Seymour, N., Stern, D., De~Breuck, C., {et~al.} 2007, ApJS, 171, 353

\bibitem[{Simpson {et~al.}(2020)Simpson, Smail, Dudzevi{\v c}i{\=u}t{\.e},
  Matsuda, Hsieh, Wang, Swinbank, Stach, An, Birkin, Ao, Bunker, Chapman, Chen,
  Coppin, Ikarashi, Ivison, Mitsuhashi, Saito, Umehata, Wang, \&
  Zhao}]{Simpson2020}
Simpson, J.~M., Smail, I., Dudzevi{\v c}i{\=u}t{\.e}, U., {et~al.} 2020, MNRAS,
  495, 3409

\bibitem[{Smail {et~al.}(1997)Smail, Ivison, \& Blain}]{Smail1997}
Smail, I., Ivison, R.~J., \& Blain, A.~W. 1997, ApJ, 490, L5

\bibitem[{Stach {et~al.}(2018)Stach, Smail, Swinbank, Simpson, Geach, An,
  Almaini, Arumugam, Blain, Chapman, Chen, Conselice, Cooke, Coppin, Dunlop,
  Farrah, Gullberg, Hartley, Ivison, Maltby, Micha{\l}owski, Scott, Simpson,
  Thomson, Wardlow, \& {van der Werf}}]{Stach2018}
Stach, S.~M., Smail, I., Swinbank, A.~M., {et~al.} 2018, ApJ, 860, 161

\bibitem[{Steidel {et~al.}(1998)Steidel, Adelberger, Dickinson, Giavalisco,
  Pettini, \& Kellogg}]{Steidel1998}
Steidel, C.~C., Adelberger, K.~L., Dickinson, M., {et~al.} 1998, ApJ, 492, 428

\bibitem[{Steidel {et~al.}(2000)Steidel, Adelberger, Shapley, Pettini,
  Dickinson, \& Giavalisco}]{Steidel2000}
Steidel, C.~C., Adelberger, K.~L., Shapley, A.~E., {et~al.} 2000, ApJ, 532, 170

\bibitem[{Stevens {et~al.}(2003)Stevens, Ivison, Dunlop, Smail, Percival,
  Hughes, R{\"o}ttgering, {van Breugel}, \& Reuland}]{Stevens2003}
Stevens, J.~A., Ivison, R.~J., Dunlop, J.~S., {et~al.} 2003, Nature, 425, 264

\bibitem[{Swinbank {et~al.}(2006)Swinbank, Chapman, Smail, Lindner, Borys,
  Blain, Ivison, \& Lewis}]{Swinbank2006}
Swinbank, A.~M., Chapman, S.~C., Smail, I., {et~al.} 2006, MNRAS, 371, 465

\bibitem[{Tanaka {et~al.}(2011)Tanaka, De~Breuck, Kurk, Taniguchi, Kodama,
  Matsuda, Packham, Zirm, Kajisawa, Ichikawa, Seymour, Stern, Stockton,
  Venemans, \& Vernet}]{Tanaka2011}
Tanaka, I., De~Breuck, C., Kurk, J.~D., {et~al.} 2011, PASJ, 63, S415

\bibitem[{Umehata {et~al.}(2019)Umehata, Fumagalli, Smail, Matsuda, Swinbank,
  Cantalupo, Sykes, Ivison, Steidel, Shapley, Vernet, Yamada, Tamura, Kubo,
  Nakanishi, Kajisawa, Hatsukade, \& Kohno}]{Umehata2019}
Umehata, H., Fumagalli, M., Smail, I., {et~al.} 2019, Science, 366, 97

\bibitem[{Umehata {et~al.}(2018)Umehata, Hatsukade, Smail, Alexander, Ivison,
  Matsuda, Tamura, Kohno, Kato, Hayatsu, Kubo, \& Ikarashi}]{Umehata2018}
Umehata, H., Hatsukade, B., Smail, I., {et~al.} 2018, PASJ, 70, 65

\bibitem[{Wang {et~al.}(2021)Wang, Hill, Chapman, Wei{\ss}, Scott,
  Apostolovski, Aravena, Archipley, B{\'e}thermin, Canning, De~Breuck, Dong,
  Everett, Gonzalez, Greve, Hayward, Hezaveh, Jarugula, Marrone, Phadke,
  Reuter, Rotermund, Spilker, \& Vieira}]{Wang2021}
Wang, G. C.~P., Hill, R., Chapman, S.~C., {et~al.} 2021, MNRAS, 508, 3754

\bibitem[{Wylezalek {et~al.}(2013{\natexlab{a}})Wylezalek, Galametz, Stern,
  Vernet, De~Breuck, Seymour, Brodwin, Eisenhardt, Gonzalez, Hatch, Jarvis,
  Rettura, Stanford, \& Stevens}]{Wylezalek2013a}
Wylezalek, D., Galametz, A., Stern, D., {et~al.} 2013{\natexlab{a}}, ApJ, 769,
  79

\bibitem[{Wylezalek {et~al.}(2013{\natexlab{b}})Wylezalek, Vernet, De~Breuck,
  Stern, Galametz, Seymour, Jarvis, Barthel, Drouart, Greve, Haas, Hatch,
  Ivison, Lehnert, Meisenheimer, Miley, Nesvadba, R{\"o}ttgering, \&
  Stevens}]{Wylezalek2013}
Wylezalek, D., Vernet, J., De~Breuck, C., {et~al.} 2013{\natexlab{b}}, MNRAS,
  428, 3206

\bibitem[{Zeballos {et~al.}(2018)Zeballos, Aretxaga, Hughes, Humphrey, Wilson,
  Austermann, Dunlop, Ezawa, Ferrusca, Hatsukade, Ivison, Kawabe, Kim, Kodama,
  Kohno, Monta{\~n}a, Nakanishi, Plionis, {S{\'a}nchez-Arg{\"u}elles}, Stevens,
  Tamura, Velazquez, \& Yun}]{Zeballos2018}
Zeballos, M., Aretxaga, I., Hughes, D.~H., {et~al.} 2018, MNRAS, 479, 4577

\bibitem[{Zhang {et~al.}(2022)Zhang, Zheng, Shi, Gao, Dannerbauer, An, Shu,
  Gao, Wang, Wang, Cai, Fan, Fang, Pan, Liu, Tan, Qin, Ren, Qiao, Wen, \&
  Liu}]{Zhang2022}
Zhang, Y., Zheng, X.~Z., Shi, D.~D., {et~al.} 2022, MNRAS, 512, 4893

\bibitem[{Zirm {et~al.}(2008)Zirm, Stanford, Postman, Overzier, Blakeslee,
  Rosati, Kurk, Pentericci, Venemans, Miley, R{\"o}ttgering, Franx, {van der
  Wel}, Demarco, \& {van Breugel}}]{Zirm2008}
Zirm, A.~W., Stanford, S.~A., Postman, M., {et~al.} 2008, ApJ, 680, 224

\bibitem[{Zwaan {et~al.}(2022)Zwaan, Ivison, Peroux, Chen, Klitsch, Hamanowicz,
  Szakacs, Weng, Biggs, \& Smail}]{Zwaan2022}
Zwaan, M., Ivison, R., Peroux, C., {et~al.} 2022, Messenger, pp. 10-13, 4 pages

\end{thebibliography}

\begin{appendix}
  \section{Source classification}\label{appendix:classification}
\begin{figure}[h]
   \centering
   \includegraphics[width=\linewidth]{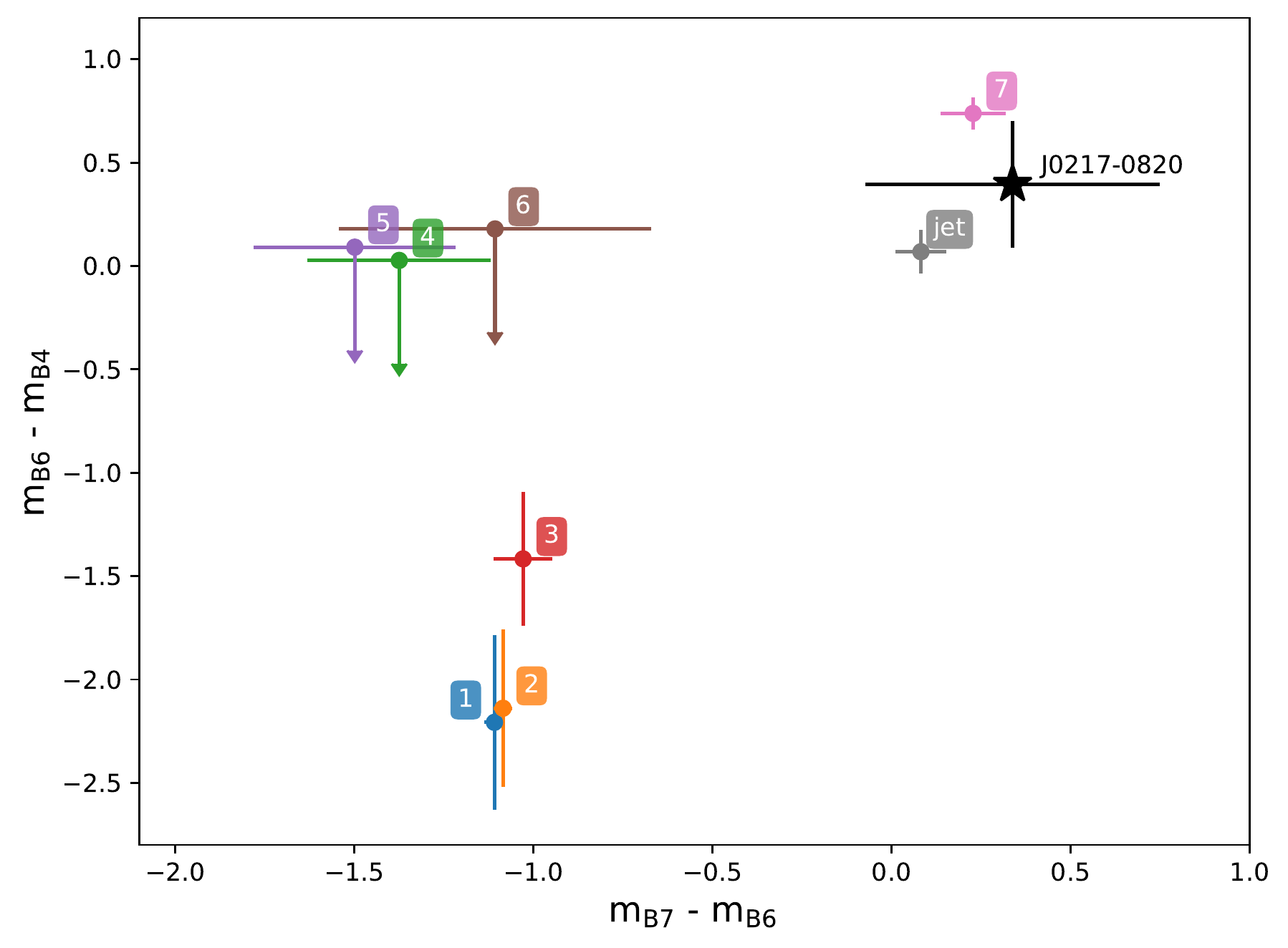}
   \caption{Submm/mm colours of the detected sources. The colours of the synchrotron radio sources are quite distinct from those of the DSFGs.}
   \label{fig:color_color}
\end{figure}

We classified the sources based on multi-band photometry from ALMA.
For thermal emission from dust, the submm/mm flux densities of DSFGs decrease with increasing wavelength; on the other hand, for synchrotron emission, their flux densities will increase.
These emission mechanisms can therefore be aptly separated by their submm/mm colours.
We constructed a colour-colour plot based on ALMA bands 4, 6, and 7 for all the sources in Fig.\,\ref{fig:color_color}.
The three bright SMGs share a similar spectral index in the ALMA bands, which is different from the radio sources, including the blazar and jet.
The other three DSFGs are only detected in bands 6 and 7, so we used their upper limits in band 4 to calculate the colours.
Although they suffer from large uncertainties, their submm colours are close to those of the SMGs.
In band 3, we did not detect the continuum emission of the DSFGs; also, the central band-3 region is contaminated by the blazar. For these reasons, we did not include the band-3 continuum measurements in Table.\,\ref{tab:continuum}.

\section{Spectral lines}
\begin{figure*}
   \centering
   \includegraphics[width=\linewidth]{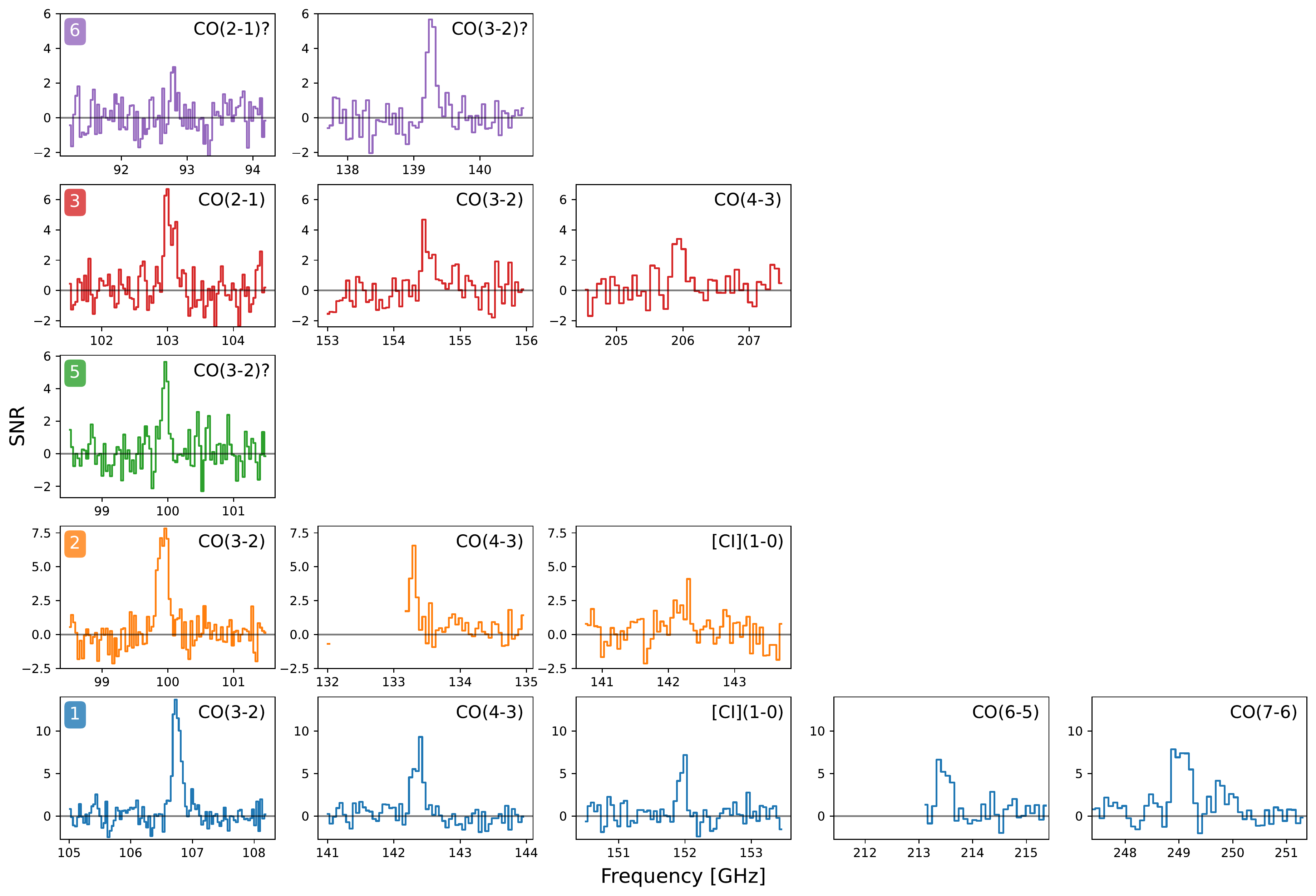}
   \caption{Zoomed-in view of the spectral lines in Fig.~\ref{fig:spectra}, with additional lines found in ALMA band 5 and band 6.}
   \label{fig:spectra_zoom}
\end{figure*}

We measured the line fluxes of all the detected emission lines in Fig.\,\ref{fig:spectra_zoom}.
We extracted the spectra using a fixed aperture, twice the size of the synthesised beam.
Since most of the sources remain unresolved or only marginally resolved in our images, this aperture gives the most robust results based on our tests.
All the emission lines detected in bands 3 and 4 have negligible continuum emission, so we fitted the extracted spectra with a single Gaussian line profile. 
For the lines detected in bands 5 and 6, we fitted the line with a Gaussian profile plus a flat continuum.
The resulting line fluxes are reported in Table\,\ref{tab:spectral_lines}.

\begin{table*}
  \centering
  \caption{Measured flux of the confirmed spectral lines.}
  \label{tab:spectral_lines}
  \begin{tabular}{ccccccccc}
    \hline\hline
    ID & CO(2--1) & CO(3--2) & CO(4--3) & CO(5--4) & CO(6--5) & [C{\sc i}](1--0) & CO(7--6) \\ 
       & (Jy\,km\,s$^{-1}$) &(Jy\,km\,s$^{-1}$) & (Jy\,km\,s$^{-1}$) &(Jy\,km\,s$^{-1}$) & (Jy\,km\,s$^{-1}$) & (Jy\,km\,s$^{-1}$) & (Jy\,km\,s$^{-1}$)\\ 
    \hline
HLW-1 & -- & 1.47$\pm$0.08 & 1.70$\pm$0.15 & -- & 2.53$\pm$0.22 & 0.71$\pm$0.12 & 1.33$\pm$0.11 \\
HLW-2 & -- & 1.44$\pm$0.10 & 1.25$\pm$0.18 & -- & -- & 1.52$\pm$0.24 & -- \\
HLW-3 & 1.06$\pm$0.20 & 1.07$\pm$0.20 & 1.20$\pm$0.20 & -- & -- & 0.61$\pm$0.04 & -- \\
HLW-5 & -- & 0.72$\pm$0.20$^{a}$ & -- & -- & -- & -- & -- \\
HLW-6 & 0.34$\pm$0.09 & 0.36$\pm$0.08 & -- & -- & -- & -- & -- \\
    \hline
  \end{tabular}
  \flushleft
  \textit{Notes:} (a) assumes the same redshift as HLW-2.
\end{table*}

\section{Simulations}\label{appendix:simulations}

Recently, numerical simulations have successfully reproduced the number counts of DSFGs over a wide range of flux densities \citep{Chen2023}. 
These simulations can thus be used to test the projection effect with various setups.
Depending on the adopted methodologies, different works have used different assumptions and approximations.
It is therefore worth carrying out these tests in the context of different types of simulations.

We adopted the semi-analytical simulation from the SHARK project \citep{Lagos2019,Lagos2020}, which has a cubic volume of $(210\,\mathrm{cMpc}\,h^{-1})^3$, and calculated the dust emission from the SED template \textsc{ProSpect}\footnote{https://github.com/asgr/ProSpect}.
We also adopted the semi-empirical simulation from \citet{Popping2020}, which plugged the observational scaling relations into \textsc{UnivereMachine} \citep{Behroozi2019} and produced a simulated cubic Universe with a volume of $(250\,\mathrm{cMpc}\,h^{-1})^3$.
The submm/mm flux densities of the galaxies are scaled from their SFR and dust mass \citep{Popping2020}.
For the hydrodynamical simulations, we used the light cones from \citet{Lovell2021}, which performed full radiative transfer simulations for all the galaxies with SFR$>20\,M\odot\mathrm{yr}^{-1}$ in the \textsc{Simba} $(100\,\mathrm{cMpc}\,h^{-1})^3$ box \citep{Dave2019a}.
This simulation is only complete for DSFGs with $S_{850}>0.25$\,mJy, similarly to the case of the faintest DSFGs found around \JJ{}.

With all these simulated light cones, we searched for DSFGs following the same observational setups.
We used the same FoV as we did in band 7 ($d=30''$). 
We randomly placed the pointing within the simulated light cone and then counted the number of DSFGs as a function of flux density in each pointing.
We repeated this process 1000 times to obtain the average number of DSFGs at each flux level and to sample their variations.
We note that the simulations did not all model the radio emission of the galaxies, so we can only compare the results for DSFGs.
We did not apply redshift constraints when searching for over-densities in the simulated light cones, as the dominant population of DSFGs lies at $z>1$ \citep{Lagos2019,Popping2020}.

\section{ALMA projects}\label{appendix:almaid}
2015.1.00456.S,
2015.1.00862.S,
2015.1.00820.S,
2016.1.00282.S,
2016.1.00434.S,
2016.1.00754.S,
2016.1.01184.S,
2016.1.01172.S,
2016.1.01262.S,
2017.1.00413.S,
2017.1.00562.S,
2017.1.01027.S,
2017.1.01492.S,
2017.1.01674.S,
2018.1.00164.S,
2018.1.00478.S,
2018.1.00490.S,
2018.1.00657.S,
2018.1.00815.S,
2018.1.00828.S,
2018.1.00922.S,
2018.1.00966.S,
2018.1.01044.S,
2018.1.01140.S,
2018.1.01188.S,
2019.1.00074.S,
2019.1.00102.S,
2019.1.00337.S,
2019.1.00363.S,
2019.1.00397.S,
2019.1.00900.S,
2019.1.01027.S,
2019.1.01329.S,
2019.1.01529.S,
2019.1.01634.L,
2021.1.00207.S,
2021.1.00666.S,
2021.1.00705.S,
2021.1.01342.S,
2021.1.01535.S,
2021.1.01650.S,
2021.1.01683.S,

\end{appendix}

\end{document}